# Accelerating Progress Towards Practical Quantum Advantage:
## The Quantum Technology Demonstration Project Roadmap

## ABSTRACT

Quantum information science and technology (QIST) is a critical and emerging technology with the potential for enormous world impact and currently invested in by over 40 nations.  To bring these large-scale investments to fruition and bridge the lower technology readiness levels (TRLs) of fundamental research at universities to the high TRLs necessary to realize the promise of practical quantum advantage accessible to industry and the public, we present a roadmap for Quantum Technology Demonstration Projects (QTDPs).  Such QTDPs, focused on intermediate TRLs, are large-scale public-private partnerships with high probability for translation from laboratory to practice. They create technology demonstrating a clear 'quantum advantage' for science breakthroughs that are user-motivated and will provide access to a broad and diverse community of scientific users. Successful implementation of a program of QTDPs will have large positive economic impacts.

## EXECUTIVE SUMMARY

Critical and emerging technologies are defined as a subset of advanced technologies that are potentially significant to national security [1].  Quantum information science and technology (QIST) is one such technology, that has made rapid progress in recent years and appears on the cusp of transforming our notions of computing, communications, and sensing. Over 40 nations have invested in large scale quantum information science and technology (QIST) program worldwide [2]. These investments have tended to focus, on the one hand, on fundamental research, and, on the other, in supporting industry from start-ups to large-scale international companies.  The concept of technology readiness levels (TRLs) 1-9 [3] encompasses the scope of these investments.  Large-scale government investment at TRLs 4-6 is presently lacking in the US and many other nations, while TRLs 7-9 is often dominated by a focus on commercial development *in lieu* of the immediately useful scientific applications.  We propose Quantum Technology Demonstration Projects (QTDPs) be created focused on TRLs 4-6 to bridge TRLs 1-3, often found at universities and national labs, to TRLs 7-9 for scientific applications currently being pursued in QIST industry.  Such QTDPs will bridge the existing gaps, achieving the current goal of large-scale industrial realizations of commercial development.  We put forth a concrete roadmap for realizing this investment.

Since the launch of the U.S. National Quantum Initiative (NQI) Act in 2018, many U.S. agencies, including the Department of Defense (DOD), Department of Energy (DOE), and National Science Foundation (NSF), have taken important strides to accelerate the maturation of QIST research and to build teams and communities of scientists and engineers working toward common goals. For some programs that are the farthest along, the time is right to follow those developments with a push to nurture and accelerate the translation from laboratory to practice having the largest scientific and economic impact. This roadmap does not attempt to advise or recommend detailed schemes for managing such a program; rather, it provides the factual bases for evaluating the merits of launching such a program.

A central goal of the development cycle is to demonstrate a 'quantum advantage' over competing classical approaches.  By quantum advantage, we refer to the ability of a quantum system to perform a useful task, including performance characteristics that are:

- Faster or better than is possible with an existing classical system (*e.g.*, Shor's algorithm can find prime numbers faster than classical algorithms); or
- Likely not possible using any existing classical system (*e.g.*, extending the baseline of an interferometric optical telescope array).

'Useful' can refer to a scientific, industrial, defense, or societal use. It is our consensus that, for practical reasons, scientific uses will be the most important in the short term, an emphasis that is aligned with the core mission of agencies focused on fundamental research.   At a more detailed level, quantum advantage should be defined according to the classes of use cases (*e.g.*, quantum computing advantage, quantum sensing advantage, quantum communication or networking advantage, *etc.*).

We have identified the following QIST projects, rank-ordered by the readiness for a major push via large-scale supported QTDPs:

1. Accelerate the quantum advantage of quantum computers by developing the full-stack operation to address scientific and engineering grand challenges;
2. Quantum simulation of high-impact scientific problems including fundamental physics, materials design, and quantum chemistry;
3. Ensembles of quantum sensors for precision measurement in the physical, biological, and medical sciences and networks of quantum sensors interconnected to quantum computers for coherent quantum signal processing;
4. Quantum networking, including entanglement distribution and teleportation across metropolitan-scale distances as an enabler for multiple applications;
5. Applications of quantum entanglement-distributions networks including clock synchronization, imaging, geodesy, secure voting, and very-long baseline interferometry (VLBI) for telescopic imaging in the visible spectrum;
6. Quantum data centers that include modular quantum computers with memories and interconnects.

Our consensus is that these projects have high probability to lead to translation from laboratory to practice and those at the top of the list are ready to begin if there is a mechanism to support them.  They will realize technologies that demonstrate a clear 'quantum advantage' and will enable scientific and engineering breakthroughs.

The success of these projects will be increased by focusing their efforts using an application-driven and user-motivated approach that engages a broad and diverse community of scientific users.  These projects will also likely benefit from having a joint solicitation by multiple government agencies, where each agency might decide to fund a distinct component of the project.

To support these projects and other emerging quantum technologies, we also envision creating a network of "Distributed Quantum Technology Hubs." Here, we envision several regional hubs that have a "quantum makers" space, with expertise and equipment required to fabricate low-volume, cross-cutting, state-of-the-art devices that will vastly increase the quantum toolbox. Users can visit the hub to work alongside the experts and be trained in the fabrication and manufacturing process and the hub can provide resources for early-stage quantum startup companies. The hubs will also sub-contract to university and industry partners to continue their development of key quantum technologies. These hubs should be informed by the quantum metrology and certification programs; in the US, these would be at the National Institute of Standards and Technology (NIST) and the Quantum Economic Development Consortium (QED-C).

**Accelerating Progress Towards Practical Quantum Advantage:**
**The Quantum Technology Demonstration Project Roadmap**

Paul Alsing*, US Air Force Research Laboratory
Phil Battle, AdvR, Inc.
Joshua C. Bienfang[#], National Institute of Standards and Technology
Tammie Borders, Quantinuum
Tina Brower-Thomas, Howard University
Lincoln D. Carr, Colorado School of Mines
Fred Chong, University of Chicago
Siamak Dadras, TOPTICA Photonics, Inc.
Brian DeMarco, University of Illinois, Urbana-Champaign
Ivan Deutsch, University of New Mexico
Eden Figueroa, Stony Brook University
Danna Freedman, Massachusetts Institute of Technology
Henry Everitt*, Army Research Laboratory
Daniel Gauthier+, The Ohio State University
Ezekiel Johnston-Halperin, The Ohio State University
Jungsang Kim, Duke University and IonQ
Mackillo Kira, University of Michigan
Prem Kumar, Northwestern University
Paul Kwiat, University of Illinois, Urbana-Champaign
John Lekki[%], National Aeronautics and Space Administration
Anjul Loiacono, Cold Quanta
Marko Loncar, Harvard University and Hyperlight
John R. Lowell, The Boeing Company
Mikhail Lukin, Harvard University
Celia Merzbacher, SRI International
Aaron Miller, Quantum Opus
Christopher Monroe, Duke University and IonQ
Johannes Pollanen, Michigan State University and EeroQ Quantum Hardware
David Pappas, Rigetti Computing
Michael Raymer+, University of Oregon
Ronald Reano, The Ohio State University
Brandon Rodenburg, The Mitre Corporation
Martin Savage, University of Washington
Thomas Searles, University of Illinois, Chicago
Jun Ye[#], JILA, National Institute of Standards and Technology, and The University of Colorado

+ Corresponding authors, e-mail: DG: gauthier.51@osu.edu; MR: raymer@uoregon.edu
* These opinions, recommendations, findings, and conclusions do not necessarily reflect the views or policies of the DoD or the United States Government
# These opinions, recommendations, findings, and conclusions do not necessarily reflect the views or policies of NIST or the United States Government

## I. Introduction

Quantum Information Science and Technology (QIST) is poised to have a profound impact on scientific discovery, innovation, and the global economy.  The impact is driven by the possibility that quantum systems display a 'quantum advantage,' operationally defined below, that will enable computations and simulations that are not possible with today's classical computers, novel communication systems secured by fundamental principles of nature, and sensors that have greatly enhanced capabilities relative to those operating on the principles of 'classical' physics.  These technologies cross traditional topical boundaries and their development requires expertise from computer science, electrical engineering, materials science, and physics amongst other fields, which is an example of convergent research and development.

Since the launch of the U.S. National Quantum Initiative (NQI) in 2018 [4] and parallel programs worldwide [2], important strides have been taken to step up and accelerate the maturation of QIST research and to build teams and communities of scientists and engineers working toward common goals. For example, the NSF has recently sponsored project scoping workshops on the topics of quantum computers for scientific discovery [5], quantum interconnects [6], and quantum simulators [7], which have been published as roadmaps.  For some QIST topics that are the farthest along the development timeline, the time is right to follow those developments with a push to nurture and accelerate achieving a practical quantum advantage and the translation from laboratory to practice, and thus achieve the largest scientific and economic impact.

### I.A. The Quantum Information Development Cycle

The typical evolution of quantum information science and technology is illustrated in Fig. 1, which loosely corresponds to technology readiness levels (TRLs).  The beginning of the process is largely curiosity-driven research that establishes the basic science concept (1), followed by the initial technology development (2) and experimental proof-of-concept demonstrations (3). For the most promising concepts having a successful demonstration, the next phase is a continuous loop of scaling-up and improving the technology (4 → 5 → 6 → 4), realizing more advanced application demonstrations (5), and demonstrating an increasing quantum advantage over classical approaches (6).  "Scaling" refers to increasing the number of quantum-coherent elements or components in a system or to increasing the distance over which quantum coherence can be maintained.  For some examples, there are continuous partial exits out of the cycle to integrate the technology into a "full stack" operation, which opens the technology to user access (8) and enables scientific applications (9), reaching the high point of the usual TRL designations 1-9. This last step might be followed by commercial development (10), with the initial commercial driver being the scientific applications.  The later stages (7-9) often happen in parallel with the next cycle in the loop (4 → 5 → 6 → 4).  It is common to associate TRLs 8-9 with commercial products in industry available to the broader public, but in the case of QIST, as with other early stage critical and emerging technologies, this is, in our view, premature. Instead, for QIST, stage 10 naturally builds on user access (8) and scientific applications (9), which lay the groundwork for broad adaption in society and commercial development (10) – classical computing followed a similar pathway in the post-World War II era.

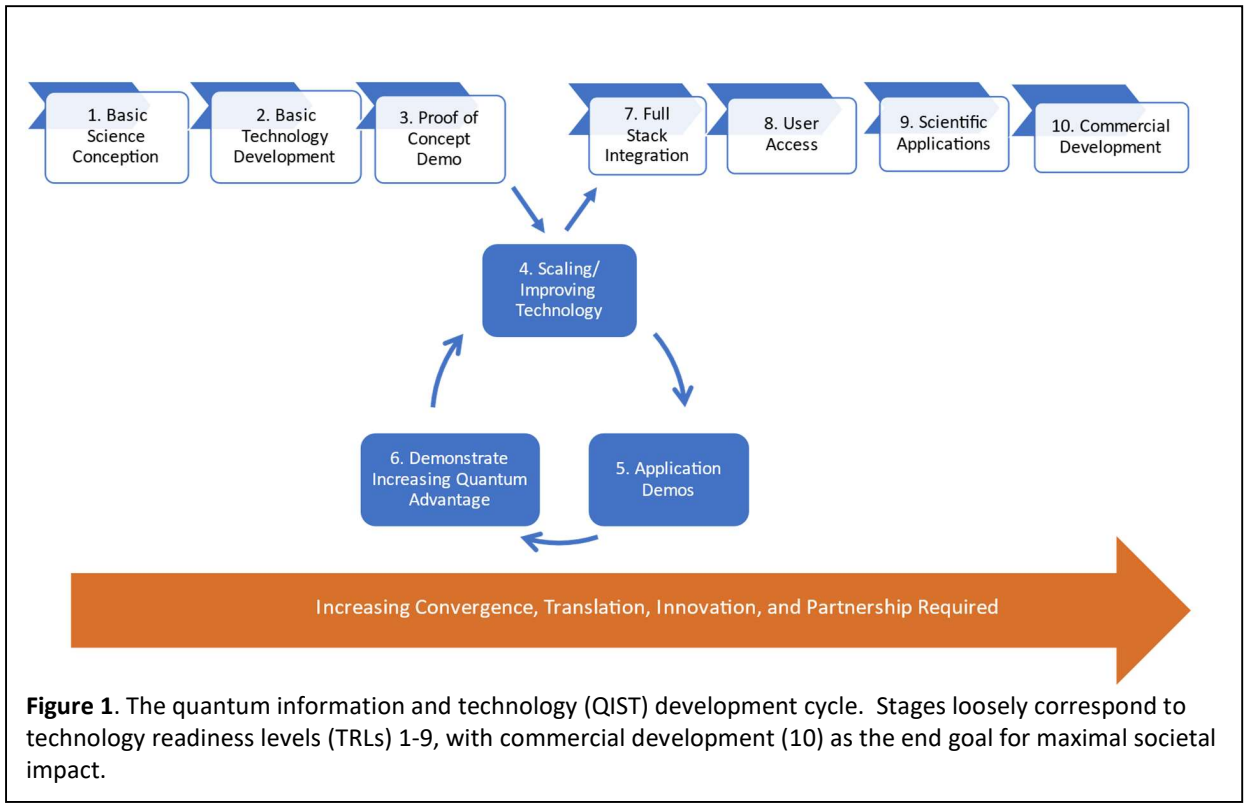

**Figure 1**. The quantum information and technology (QIST) development cycle. Stages loosely correspond to technology readiness levels (TRLs) 1-9, with commercial development (10) as the end goal for maximal societal impact.

This development cycle is currently hindered by the lack of streamlined engineering capability in building quantum components, devices, and systems, as well as an insufficient workforce in quantum systems engineering, scalable device architectures, quantum algorithm development, and the inability of conventional research groups to encompass a necessary full-stack system-level approach that integrates the physical sciences, engineering fields, computer science, and potential users from broad disciplines. Getting beyond these barriers will require increasing convergence involving researchers from multiple areas of expertise; translation of the technology from the lab to the field; scientific, engineering, and manufacturing innovation; and partnership with a wide range of technologists located in universities, government agencies and labs, and industry.

### I.B. Demonstrating a Quantum Advantage

To validate a technology, one milestone in the cycle is demonstrating a 'quantum advantage' over competing classical approaches (6 in Fig. 1). Here, we agree that a quantum advantage refers to the ability of a quantum system to perform a useful task, including performance characteristics that are:

- Faster or better than is possible with an existing classical system (*e.g*, Shor's algorithm); or

- Likely not possible or practical using any existing classical system (*e.g.*, a quantum teleportation network or extending the baseline of an optical interferometric telescope array).

'Useful' can refer to a scientific, industrial, defense, or societal use. We believe that, for practical reasons, scientific uses will be the most important in the short term, an emphasis that is aligned with the core mission of fundamental research-focused agencies.  Moreover, it is our view that scientific use cases of quantum technology will catalyze commercial development.

At a more detailed level, quantum advantage should be defined according to the classes of use cases (*e.g.*, quantum computing/simulation advantage, quantum sensing advantage, quantum communication or networking advantage, *etc.*).  It should be the responsibility of a project proposer to clearly define the quantum advantage being targeted by their proposed project. As such, there should be flexibility in evaluating a proposer's particular definition or criterion.

### *I.C. Quantum Technology Demonstration Project Concept*

We agree that the QIST community is capable of and ready to stand up large-scale ***Quantum Technology Demonstration Projects*** (QTDPs). By 'large-scale,' we mean roughly an order of magnitude larger than previous fundamental-research-focused projects aimed at developing a particular quantum technology. In the US the new NSF Technology, Innovation, and Partnership (TIP) Directorate, for example, has or could have the appropriate set of capabilities to support an overarching program to support this effort, previously referred to in NSF documents as the *NSF National Virtual Laboratory in QIST*. This roadmap does not attempt to advise or recommend detailed schemes for managing such a program; rather, it provides the factual bases for evaluating the merits of launching such a program.

Key to the success of demonstrator projects is that they represent the broad quantum portfolio of computing, simulation, sensing, and communication.  It is especially important that the QTDPs have a majority representation from the engineering, applied science, and computer science communities, who are needed to scale up what are often one-off research-focused devices.  Having strong activities in multiple areas will 'quantum activate' researchers in a variety of traditional fields to contribute their knowledge and expertise to the projects. Advances in technologies from one project will have an impact on the others. Finally, multiple areas of activity will attract broader participation in the field by providing many career pathways.

## II. Quantum Technology Demonstration Projects

To arrive at a selection of topics that are primed for launching a QTDP, we considered the current state of each suggested topic along the development cycle shown in Fig. 1. We reached a consensus that topics that are now ready to scale up to a QTDP have already gone through at least one scaling loop (4 → 5 → 6 → 4) and have already demonstrated the ability to spin out the technology towards scientific discovery (7-9) or commercial application (10).  Here, we assume that some user applications (8) are solving known problems, whereas others will be focused on scientific discovery (9), and that both uses will motivate continued commercialization. Other topical areas may have also followed this path but are just entering the scaling loop (4 → 5 → 6 → 4) and hence these topical areas can benefit from sub-community-wide planning to focus on a smaller number of use-inspired and application-driven projects. Less developed but highly impactful areas will need additional focused research and development in parallel with sub-community-wide planning.

Another constraint is that in most cases the QTDPs will need to be led by university teams, although they can and probably should involve industry partners. Some areas of QIST already have substantial industrial investment and human resources focused on specific platforms, such as major information technology companies developing superconducting qubit quantum processors, and hence a university-led QTDP might not be competitive in scaling up to full-stack development. On the other hand, a university-led QTDP would have the advantage of not being constrained by revenue and profit considerations and could direct the technology outcomes toward scientific discovery. It is up to the eventual project teams to specify their current state on the development timeline and the scientific discovery potential of their proposed technology.

The QIST topics, rank-ordered by the readiness for a major push via large-scale QTDPs involving public-private partnerships, are illustrated in Fig. 2, and are described in greater detail in the sub-sections below. These QTDPs have a high probability to lead to translation from laboratory to practice, and those at the top of Fig. 2 are ready to begin immediately if there is a mechanism to support them. They have high potential to realize technologies that demonstrate a clear quantum advantage and will enable near-future scientific and engineering breakthroughs.

Other topics in Fig. 2 likely need additional team building and topic focusing to be prepared for standing up a QTDP. We anticipate that a grant-supported planning activity of one year is enough time for some topics, whereas some might require a two-year planning phase as indicated by the shaded bars.

There are several other topics not appearing in Fig. 2 that should continue to be supported by existing mechanisms, such as the 13 DOD, DOE, and NSF quantum centers in the US. Importantly, multiple efforts should be supported that are aimed directly at removing identified bottlenecks to move such topics closer to being ready for a large-scale demonstration project. For example, high-rate high-efficiency quantum repeaters for long distance entanglement distribution are still missing, although many concepts for building such systems are known; these could be pursued by teams that are aiming to stand up a QTDP in the topic of long-distance networking.

### II.A QTDP-ready topics

The success of the QTDPs will be increased by focusing their efforts using an application-driven and user-motivated approach that engages a broad and diverse community of scientific users throughout the country. In this section, we offer greater detail on the scope of the possible projects following the ordering in Fig. 2 except for the distributed quantum technology hubs, which are described in Sec. II.B. As mentioned previously, the first two topics (*i.e.*, those related to quantum computation and quantum simulation) in the figure could be funded immediately.

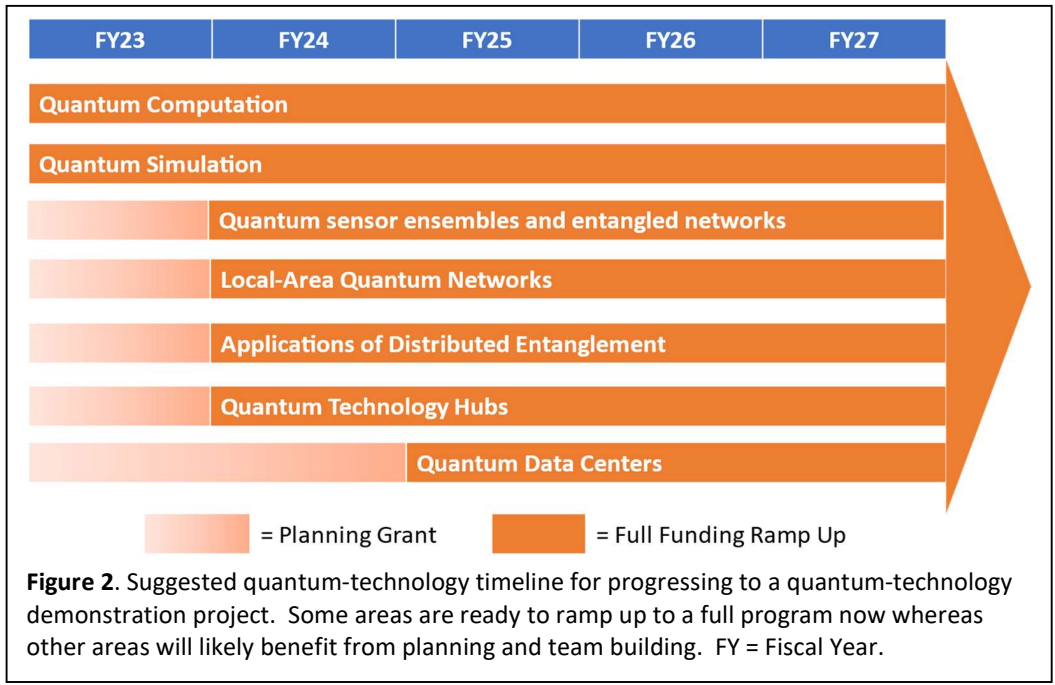

**Figure 2**. Suggested quantum-technology timeline for progressing to a quantum-technology demonstration project. Some areas are ready to ramp up to a full program now whereas other areas will likely benefit from planning and team building. FY = Fiscal Year.

We envision that the QTDPs will be distributed and incorporate expertise across multiple domains (convergence). Teams will include researchers across the spectrum of activities in phases (7 to 9) in Fig. 1, including an active user base who would develop new applications designed for supporting the quantum hardware and who would use the technology for scientific discovery. For topics that have a classical counterpart, such as quantum computation, the teams will include classical algorithm developers who continuously improve the state-of-the-art of classical algorithms to help define the boundary of a quantum advantage. Finally, we strongly believe that all QTDPs will operate in a fully transparent and open-source approach so that everyone can benefit from their developments, and that users will have the ability to fully control the hardware so that novel algorithms can be explored and developed that take advantage of the hardware characteristics.

### *II.A.i Quantum computation*

The demonstration of a universal, programable quantum computer capable of running arbitrary quantum algorithms with reliable output is arguably the grand challenge of QIST, which significantly overlaps with other quantum technology goals, including quantum simulation and quantum networking, as discussed in the sections below. There are a variety of architectures for universal quantum computing, including universal gate families, measurement/feedforward, and direct Hamiltonian control, illustrated in Fig. 3. All are computationally equivalent, in principle, to the hypothetical quantum Turing machine, a digital architecture based on qubits, a discrete set of universal quantum gates, and measurements in a computational basis. The rapid development of gate-based quantum computers was ignited by Shor's factoring algorithm. In the US, for example, this was accelerated with research and development (R&D) at universities fueled by support for basic research NSF and at national laboratories funded in part by defense and intelligence agencies (an early example is [8]). Today there is major investment in development of quantum computers in companies across the industrial landscape including large technology companies, government contractors, startups, and public-private partnerships, spurred by the National Quantum Initiative Act. Other nations have followed patterns of investment suitable to their own grant and TRL ecosystems, but a common factor is the challenge in bridging TRLs (1)-(3) to (7)-(9) to achieve commercial development (10).

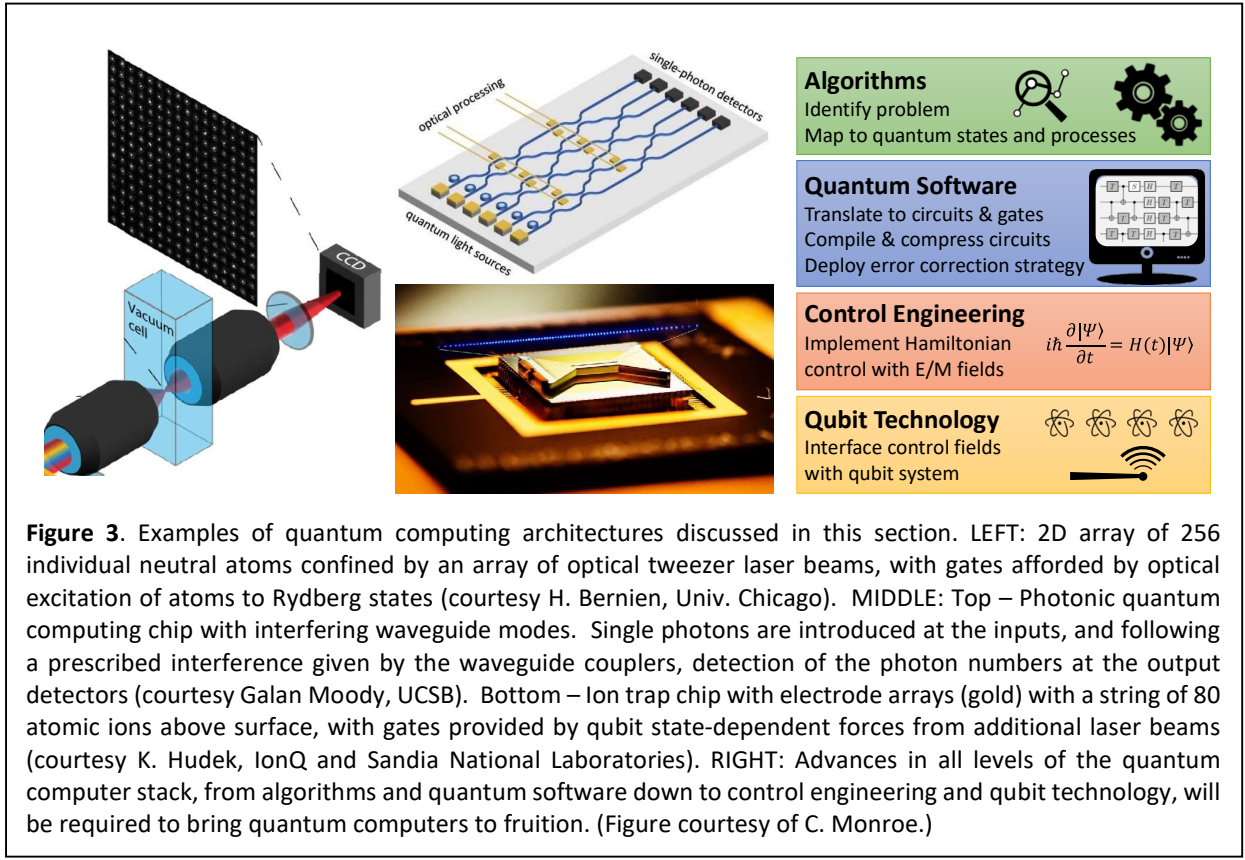

**Figure 3**. Examples of quantum computing architectures discussed in this section. LEFT: 2D array of 256 individual neutral atoms confined by an array of optical tweezer laser beams, with gates afforded by optical excitation of atoms to Rydberg states (courtesy H. Bernien, Univ. Chicago). MIDDLE: Top – Photonic quantum computing chip with interfering waveguide modes. Single photons are introduced at the inputs, and following a prescribed interference by the waveguide couplers, detection of the photon numbers at the output detectors (courtesy Galan Moody, UCSB). Bottom – Ion trap chip with electrode arrays (gold) with a string of 80 atomic ions above surface, with gates provided by qubit state-dependent forces from additional laser beams (courtesy K. Hudek, IonQ and Sandia National Laboratories). RIGHT: Advances in all levels of the quantum computer stack, from algorithms and quantum software down to control engineering and qubit technology, will be required to bring quantum computers to fruition. (Figure courtesy of C. Monroe.)

The creation of a programable quantum computer that can demonstrate a quantum advantage for scientific applications by outperforming classical high-performance computers would be a major milestone for QIST, where the T refers especially to the need for participation from engineering. One class of scientific problems that can benefit from quantum computing that lies beyond the capabilities of classical computing is found in the fields of nuclear and particle physics, such as modeling the real-time dynamics of the Standard Model, including baryogenesis, inelastic electroweak-nuclear reactions, transport properties of matter under extreme conditions, and neutrino flavor dynamics in astrophysical environments. Quantum computation is well-suited to simulate the essential elements of supernova physics such as coherent neutrino flavor dynamics and the simulations of large lattices of neutrinos, with the inclusion of incoherent processes and thermal distributions. Quantum simulations of quantum chromodynamics will impact nuclear and particle physics through simulations of dynamical quantities such as responses to external probes and the determinations of viscosities and transport properties in closed and open environments. A current experimental program in fundamental symmetries is geared toward measuring or constraining the matter-antimatter asymmetry that takes place in the earliest moments of the universe and requires physics beyond the Standard Model. Quantum computing resources will also likely be required to solve the general quantum many-body problem, and a broad-class of open questions in condensed matter [9].

Achieving quantum advantage in computing will require a tightly integrated team addressing both the hardware and software components that constitute the full stack of scientific and engineering challenges.  Chief amongst these is overcoming the deleterious effects of control errors and decoherence through methods ranging from error mitigation and error correction to eventual fully fault-tolerant operation.  A QTDP will accelerate progress towards this goal with an integrated team focused on co-design with tailored error-mitigation and error-correction strategies.  This co-design approach extends from a particular hardware platform all the way up to algorithm development, be it for general purpose applications such as optimization, or special-purpose applications such as quantum chemistry and quantum simulation of condensed matter and quantum fields, as discussed above.  Other key components include benchmarking at the hardware level to characterize error channels and verify performance, and the development of a benchmarking suite for algorithmic performance of scientific applications.  The combination of hardware and software solutions from teams of scientists, engineers, and end users will be essential in an interdisciplinary effort.

While quantum computing is steadily progressing from a purely academic discipline towards implementations with the prospects highlighted above, the nature of traditional laboratory setups does not readily offer itself to scaling up system sizes or allow for applications outside of laboratory-grade environments. Transitioning from lab-based, proof-of-concept demonstrations to robust, integrated realizations of quantum computing hardware is an important step to materialize those prospects. This transition requires overcoming challenges in engineering and integration without sacrificing the state-of-the-art performance of laboratory implementations (see, for example, [10]). To this end, it is important for the QTDPs to include partnerships with quantum computing industries and adjacent quantum enabling technologies in the quantum supply chain such as lasers, control electronics, cryogenics, semiconductors, A computer architecture, etc.  The NSF QTDP quantum computing projects can leverage their industry partners to synergistically develop robust and compact demonstrators, train quantum workforce, license their technology, and mature their technology to high TRLs.

The landscape of quantum computing platforms continues to evolve, with the most promising candidates emerging.  Several large companies have made large investments in technologies based on superconducting qubits and circuits and research and development continues to scale these systems to create useful quantum computers. Platforms based on individual atoms (in free space or in solid substrates), electrons or collections of photons are also under development in research laboratories and startup companies. These qubits are natural and replicable qubits, and the challenges towards a scalable quantum computer reside less in physics breakthroughs and more in systems engineering.  Architectures based on trapped atomic ions with dense connectivity have demonstrated the highest fidelity operations and are viewed by many as the leading contender for scalable universal quantum computing.  In addition, arrays of optically trapped neutral atoms, with a backbone like trapped-ion systems, have emerged with new demonstrations of high-fidelity gates and scalable architectures.  Trapped electron systems, while earlier in the development timeline, have recently been demonstrated as novel "natural" qubits [11] that admit to precision control [12], [13] and the promise of high-fidelity gates in architectures similar to trapped ions [12]. Photonic platforms based on discrete variable (qubit, qutrit, *etc*.) and continuous variable (wave-like) encodings are also in the race and the basis of competitive startups.

The approaches based on ions, atoms and photons are likely to be those that will be most impacted by an NSF QTDP given the currently unfulfilled needed investments for large interdisciplinary science and engineering teams. One powerful model for such a QTDP will include full transparency to external users of the complete software/hardware stack, a feature that is unlikely to be available to external users of commercial quantum computers. Such transparency will benefit both the external users and the internal scientists constructing the hardware and carrying out research within the QTDP.

### *II.A.ii Quantum simulation*

Quantum simulators (QS) bridge between specialized quantum experiments to more general-purpose programmable quantum devices.  Beyond-classical calculations on quantum simulators arguably began circa 2002 and include two landmark results: the dynamics of quantum phase transitions [14]; and the paired-fermion to boson transition, called the Bardeen-Cooper-Schrieffer (BCS) to Bose-Einstein condensate (BEC) crossover [15], [16].  Since then, quantum simulations of a great variety of physical platforms have helped unravel far-from-equilibrium quantum dynamics, a topic previously nearly inaccessible to both theory and experiment, demonstrating effects as foundational as a new statistical ensemble [17] and new non-thermodynamic phases of matter [18].  They have proved a fertile ground for exploring nano-thermodynamics [19], [20], an essential understanding as the transistor size in semiconductor chips shrinks down below a few nanometers.  And perhaps above all they have been used to explore many quantum phases of matter, for example the long-sought-after phase diagram underlying high-temperature superconductors.

Quantum simulator architectures include cold and ultracold molecules, color centers, dopants in semiconductors, gate-defined quantum dots, photons in nanostructures, photons and atoms in cavities, Rydberg atom arrays, superconducting quantum circuits, trapped atomic ions, ultracold neutral atoms, van der Waals heterostructures, Moiré materials, excitons, and many more [7].  This wide variety of platforms is the same set used to construct current and projected quantum computers.  In fact, quantum simulators are a non-universal quantum computer run in continuous, rather than discrete, time.  That is, a quantum simulator is described by a continuous-time mathematical model (*i.e.*, Hamiltonian), in contrast to a gate-based quantum computer, which is described by quantum circuits in discrete time like classical computers.  For this reason, quantum simulators are sometimes called "analog" quantum computers.  In practice, they have varying degrees of programmability but go beyond one-off quantum experiments.  We note that there exist gate-based quantum simulators that are not universal machines; these devices are also considered within the scope of quantum simulators.

Like other areas of quantum information processing, most of the academic research in the quantum simulation area has been conducted using small-scale, expert-operated laboratory setups. To shift the approach from such proof-of-principle implementations towards integrated systems that can effectively realize a quantum advantage and be made available to a wide range of users, a synergistic integration of the academic research groups and the quantum enabling industries will be pivotal. The efforts need to be focused on co-design addressing both the theoretical and hardware components that constitute the full range of scientific and engineering challenges.

Close collaboration with industry will enable engineered solutions that are robust and ready for use by external researchers. Several commercial providers have already created, or are near completion of, structured cloud-access to programmable quantum simulator (PQS) platforms, extending access to groups of users that otherwise would not have the resource or expertise to apply these techniques.  Similar developments are under way across a range of different physical platforms including quantum-gas microscopes in optical-lattice simulators, trapped-ion systems, as well as arrays of superconducting qubits.

A recent development in the field that could be matured via a QTDP is the advent of PQS's. An example involves systems composed of independently translatable optical tweezers, each deterministically holding a single cold atom programmatically driven by pulses of laser light into Rydberg states [21], [22], [23], [24], [25].  These systems have already provided new insights in many-body condensed-matter system [26], [27] and large-scale quantum information systems [26], [28], [29].  Another example is a programmable all-photonic system performing boson sampling as a demonstration of quantum advantage [30].  PQSs show promise for a wide range of simulation problems in materials research and high-energy physics, with known simulation recipes ranging from new strongly correlated matter to lattice-gauge physics.  And, with new developments in using dynamic architectures [26], the roadmap for extending PQSs into gate-based architectures and even error-corrected quantum computing devices appears feasible.

A key area of exploration is complex chemistry, where quantum simulators can provide direct insight into chemical processes, e.g., simulating open quantum systems to unlock artificial photosynthesis [31], [32].  Quantum simulators also provide a platform for exploring fundamental physics, from mapping out the quantum chromodynamic (QCD) phase diagram for the universe [33], to exploring the interplay between quantum physics and classical gravity in nanomechanics experiments [34].  Past successes in this area include holographic duality, tying together ultracold atom-based quantum simulators, quark-gluon plasma experiments, and practical experimental predictions from an offshoot of string theory [35].

More generally, quantum simulators can have immediate impact in exploring new directions of quantum computing itself, towards near-term quantum advantage.  Gate-based quantum computing has, so far, mainly focused on avoiding decoherence and using unitary gates, although non-unitary gates provide an exciting direction for future dissipation-based quantum computing in which controlling qubit relaxation is used as a resource [36], [37]. Quantum simulators have likewise mainly focused on closed quantum systems, but recently dissipative quantum phase transitions have received significant attention [38] and thus the study of open quantum simulators can have broad impact in near-term quantum advantage across chemistry, physics, and computer science.

We also note that a systematic program of quantum simulator applications toward practical quantum advantage has the potential to provide a "quantum materials genome initiative" database, like databases that density functional theory (DFT) plus machine learning have provided for classical materials in the US Materials Genome Initiative [39].

### II.A.iii Quantum sensors

Quantum sensors exploit quantum phenomena for precise measurement of quantities like time, acceleration, gravity, or electro-magnetic fields. Current quantum sensors leverage single-particle effects in which each atom (or other quantum system such as a molecule, solid-state defect, *etc*.) functions as its own quantum-limited sensor, and an ensemble of particles can be measured in parallel to improve the signal-to-noise ratio (SNR) of the aggregate sensor. We are currently witnessing the translation of these sensors into other fields and the area of quantum sensing has the potential for explosive growth over the coming years. This growth builds upon precise sensing techniques that have been developed using devices such as an anionic nitrogen vacancy pair defects (NV) in diamond [40] and has potential for disruptive impact across a range of precision measurement.

We envision a phased approach to the advancement and translation of sensing technologies, with continued development of ensembles of single quantum-limited quantum sensors, with examples given below, and transitioning to arrays of entangled quantum sensors that reach the Heisenberg-uncertainty-measurement limit discussed later in this section. We envision that a QTDP may start with a focus on quantum-limited single-sensor development with strong connection to industry partners and end users from multiple disciplines, then broaden to develop networks of sensors. There is also the possibility that more than a single QTDP will be needed to realize the full potential of quantum sensors. For example, one could be focused on developing quantum-limited single-sensor technologies while another works to developed networks of entangled quantum sensors.

*Quantum Sensor Ensembles*: Focusing first on ensembles of single sensors will require uniting design tools across materials science, chemistry, and biochemistry with physics and engineering approaches to quantum control, illustrated in Fig. 4. A similar extended and interdisciplinary development cycle was required for classical biosensors, such as green fluorescent protein (GFP) or clustered regularly interspaced short palindromic repeats (CRISPR). Further, the analyte specificity will require coordinating these development efforts with end users and industry. This requires a unique co-design approach to QIST, where multiple disparate targets are pursued by teams with similar skill sets, and within a common development framework, towards a range of products. By harnessing the power of quantum-limited sensors that exploit quantum coherence to enhance sensitivity to an analyte, our community can impact areas ranging from drug design and biomedical diagnostics to geo-magnetic field detection, as well as mapping the magnetic and electrical impulses in the brain.

Translating current quantum sensors into diverse environments will require an interdisciplinary effort. This effort will need to bring in designers of quantum sensors, end users of sensors in areas ranging from biology and medicine to geochemistry, and a collaborative community of scientists and engineers working in complex systems ranging from soft condensed matter to chemistry. The aim is to build the next generation of fundamentally new tools for biology, neuroscience, geophysics, and chemistry by targeting a new generation of quantum sensors tailored to specific scientific questions. Bridging the gap between these fields will require a dedicated effort of researchers who can design, functionalize, and implement new sensors.

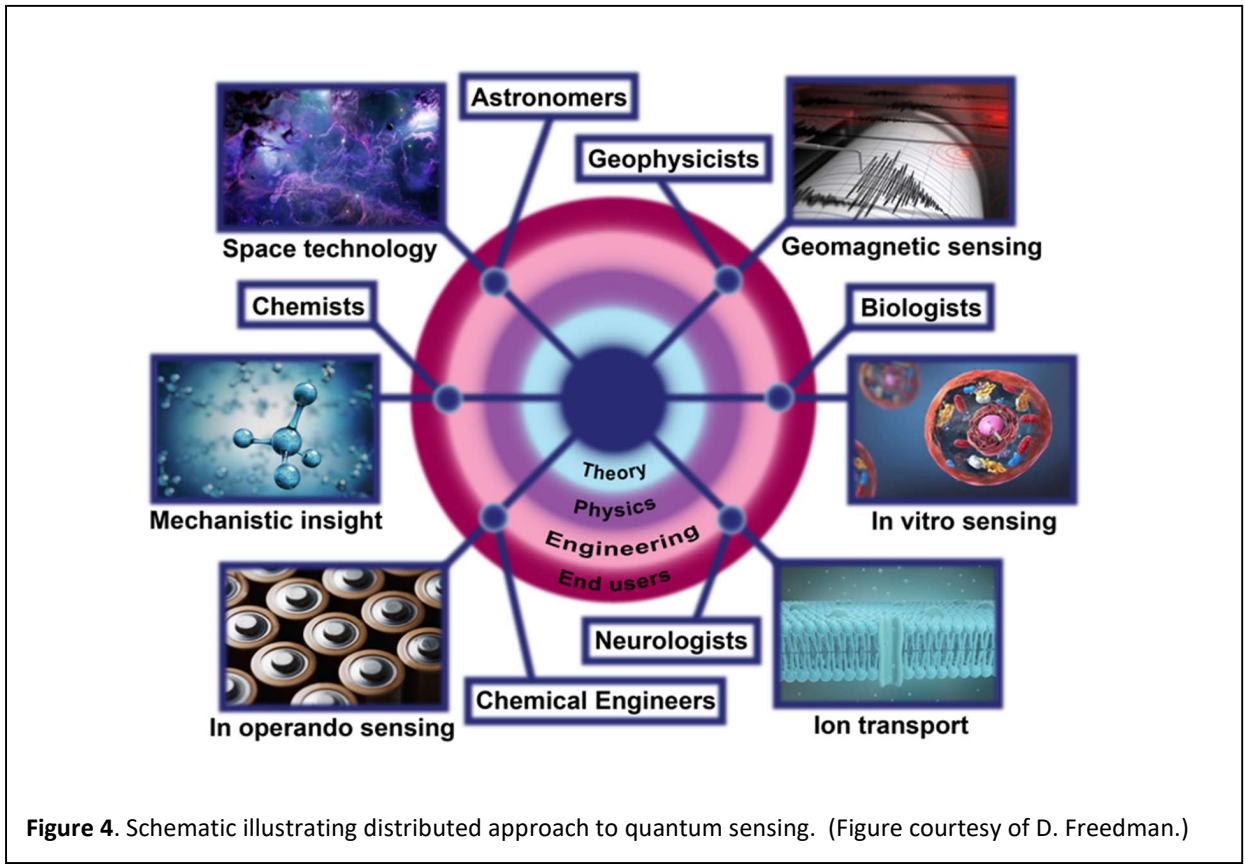

**Figure 4**. Schematic illustrating distributed approach to quantum sensing. (Figure courtesy of D. Freedman.)

As an example, targeting sensors for specific biological domains akin to double electron-electron resonance (DEER) spin labels or green fluorescent protein (GFP) is one of the first clear steps towards the quantum GFP analog. Initial research on the physics and engineering side would focus on understanding the key questions being addressed in a biological system. A sensor would need to be designed to read out magnetic signals at a single-molecule level and integrated into a biological system.

Another example is extending nuclear magnetic resonance (NMR) spectroscopy and magnetic resonance imaging (MRI) to micro- and nanoscale samples, which would enable chemical analysis of mass-limited samples that are expensive or difficult to synthesize and allow for detailed studies of cell structure and function, with applications in metabolomics and disease diagnosis. This would make single protein detection possible, opening the possibility of determining the structure of functional membrane proteins under near-physiological conditions and studying their dynamics. Membrane proteins are the targets of more than half of the Food and Drug Administration-approved drugs, and the real-time study of protein-molecule binding would facilitate drug discovery. While proof-of-concept experiments demonstrating detection of magnetic signals from individual molecules as well as NMR and MRI at the sub-cellular level have already been carried out (see, for example, [41]), integrating each of these components is insurmountable without convergent co-design between end users and quantum sensor developers.

***Quantum Sensor Networks***: In addition to individual sensors, advantage can also be achieved by leveraging integrated networks consisting of several or many quantum-entangled sensing devices.    Currently, it has been shown that such entanglement leads to sensitivity gain in quantum sensors and we are just beginning to witness true quantum advantage operating at the state-of-the-art measurement frontier [42]. The first step is to implement concrete, entanglement-enhanced quantum sensors working beyond the current state of the art. Entanglement between these constituent quantum sensors then allows for a further improvement in overall sensitivity, up to the theoretical maximum square-root improvement over the non-entangled standard limit [40]. When this entanglement is shared over a distributed network of quantum sensors, these precision enhancements can be used to provide enhanced sensitivity to targeted differential physical effects across the network or allow highly efficient intercomparison for synchronization and communication purposes [43], [44], [45].

Another instance of quantum sensor networks is the possibility to link multiple telescopes using photonic entanglement to form a large coherent array that can provide vastly increased imaging resolution in the optical spectrum, as discussed further in Section *II.A.v*.

The ability to network quantum sensors with quantum computers likewise is expected to provide broad quantum advantage. Even with near-term intermediate-scale quantum (NISQ) computers, the use of quantum computers to process quantum data such as from quantum sensors is expected to provide quantum advantage for applications related to classification [46]. For certain specific cases, such as those involving identifying quantum states or physical processes, quantum advantage is known to scale exponentially relative to classical computational methods [47], [48], [49].

### II.A.iv Local-area quantum networks

The construction of local-area quantum networks that use distributed entanglement and teleportation across building, campus or metropolitan distances will facilitate the development of hardware, software, and networking protocols and serve as an ideal enabler for testing grounds for multiple applications.  These local-area networks will enable researchers to investigate network architectures, assess network tolerance to heterogenous transduction methods, develop hardware capable of being deployed in non-laboratory environments, automating network resource allocation, real-time handling of node availability, and assess the need for network traffic management.  Each of these activities goes well beyond quantum key distribution, which is often viewed as a technical precursor to an entanglement-distribution network.

For these reasons, quantum networking is capturing the interest of a vast community across science, engineering, industry, and national security [50]. Many experts recognize that building first instances of quantum-enhanced communication networks is one of the highest priorities in QIST [51]. From financial trading to personal data handling to power grid security, quantum communication could influence many aspects of economy and society. This large relevance has already been recognized by many countries across the world, which have invested in concerted efforts to construct city-wide and inter-country quantum networks.

Today's quantum networking experiments rely on advanced quantum technology with laboratory-level functionality and performance. Among these network devices we include quantum optical sources, quantum memories with efficient input-output optical interfaces, quantum switches, signal multiplexing systems, and transducers from optical to telecommunications regimes. Broadly speaking, all these key quantum network components have yet to be run operationally in a full metropolitan network configuration. This goal will require overcoming critical challenges including achieving cascaded operation and connectivity, developing unifying operational properties, achieving high-repetition rates of entanglement distribution and demonstrating memory-assisted entanglement swapping and distillation. Figure 5 shows a concept of a metropolitan multi-node quantum network, assisted by quantum memories and switches.

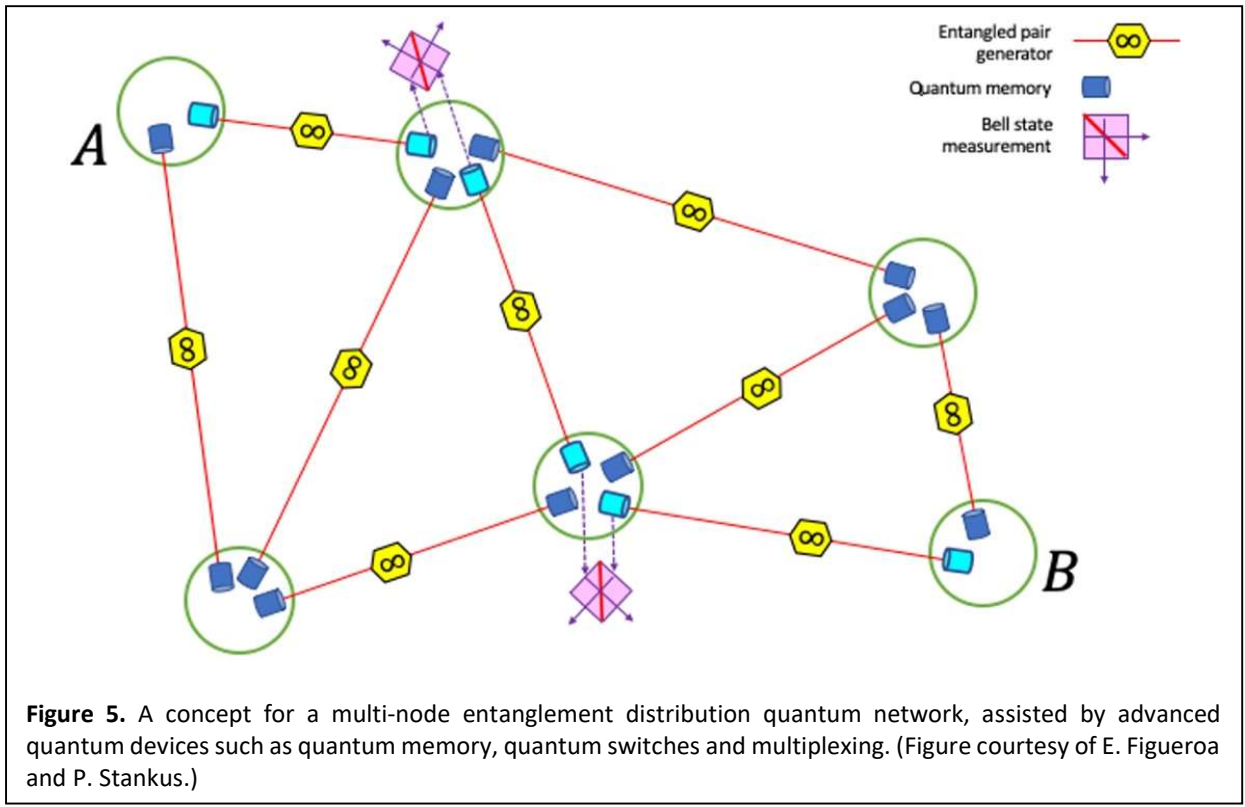

**Figure 5.** A concept for a multi-node entanglement distribution quantum network, assisted by advanced quantum devices such as quantum memory, quantum switches and multiplexing. (Figure courtesy of E. Figueroa and P. Stankus.)

To accelerate progress towards practical quantum advantage using elementary quantum networks, we suggest developing a National Scientific Quantum Network, which will support basic science applications and development using entanglement distribution across a city-wide scale, to be scaled to longer distances eventually; we note that China and the EU have both begun such developments already, to name two examples [52]. The testbed should develop and incorporate the necessary quantum technology and commonality of techniques needed to demonstrate quantum advantage in different areas of quantum communication, ranging from distributed quantum sensing to networked quantum computing. Executing this vision will require immersing and converging multi-disciplinary research teams, including quantum technology experts, fundamental material scientists and device integration experts. Of paramount relevance will be the creation of synergies with teams of quantum aware and quantum proficient engineers [53] in network-relevant areas such as computer science, network engineering and optical communication.

The current state of the art in long-distance quantum networks suggests five fundamental capabilities that can be developed and tested in an envisioned National Scientific Quantum Network. These capabilities should evolve synchronously to form a multi-purpose quantum network capable of demonstrating several first quantum communication advantage experiments. Five key capabilities and their associated demonstrations of success are as follows.

1) Quantum entanglement distribution at telecom wavelengths using already existing telecom infrastructure [54] and coexistence with classical data transmission. This should include the development of ps-level timing and synchronization and feedback mechanisms to maintain entanglement coherence in various degrees-of-freedom, such as polarization and time-bin, among inter-city nodes. *Key demonstration: Teleportation of qubits across a metropolitan area*.

2) Long-distance entanglement distribution using multiplexing in several degrees of freedom (*e.g.*, frequency or spatial mode), allowing for distribution of entanglement among many-users networks [55]. *Key demonstration: Creation of multiple pairs of entangled nodes over a complex network topology.*

3) Development of high-repetition rate quantum entanglement sources that are, by-design, compatible with more advanced quantum devices, such as quantum memories and quantum gates [56]. This will allow continuous upgrades to the testbed. *Key demonstration: storage of photonic entanglement in quantum memories located in a metro area.*

4) First demonstration of metropolitan one-hop quantum repeaters using the concepts of quantum-memory heralding and entanglement swapping. This requires the creation of several matter-matter entanglement links [57], [58] and synchronized Bell-state measurements within a metro area. *Key demonstration: quantum repeater entanglement distribution rates surpassing simple direct transmission distribution rates.*

5) Demonstration of photon-photon gate-assisted [59] multi-party entanglement creation and Bell-state measurements [60] in a metropolitan quantum repeater network. This will allow the development of error correction techniques for entanglement distribution experiments. *Key demonstration: Teleportation over non-neighboring nodes, using gate-assisted Bell-state measurements*.

Additionally, we envision that such a Scientific Quantum Network will serve as a testbed of new concepts, such as that of a quantum-enabling communication network, in which all the control and management of the quantum communication advantage demonstrations are performed using the classical internet.  Importantly, this Scientific Quantum Network should be operated using a well-defined legal framework in which universities, industry and national laboratories can work together in delivering critical experimental demonstrations.

### II.A.v Applications of quantum entanglement distribution networks

There are already numerous identified applications of quantum entanglement-distribution networks spanning a range of distance scales, including uses for quantum communication, computation, and sensing. We anticipate that more applications will be identified as quantum entanglement distribution networks mature.  While simple quantum key distribution is one of the most well-known applications of quantum networks, distributed quantum entanglement enables other, and likely more impactful, communication advantages.  For instance, by providing stronger-than-classical coordination between distributed parties, entanglement-assisted communication can lead to more efficient means of sending messages across multi-terminal networks [61]. In addition, quantum networks offer levels of security and trustworthiness that surpass what is classically achievable.  For example, spatially separated databases, such as those storing medical records or other personal information, can be checked efficiently for matching data using quantum-encryption schemes that leak no information about the data itself, a protocol known as 'quantum fingerprinting' [62], [63].  Even for relatively small computing applications, various secure multiparty computations [64], [65], [66] could be extremely useful, e.g., determining the rank ordering of multiple investors without any investor learning the precise financial value of any of the others.

A somewhat different application involves the use of distributed quantum states to verify the location of someone on a network (quantum position verification [67]); this could be particularly important, e.g., when remote parties using blind quantum computing can run any algorithm on a processor without the computer owners knowing what is being simulated — quantum position verification could at least certify that the user is in fact programming from within a secure location.  The blind quantum programming is itself a prime example of a distributed quantum network application [68], [69],  and is the most likely way that many users will have equitable access to share quantum-enabled capabilities, by providing secure remote access to quantum computers or sensor networks. The development of hybrid computing – combining both quantum and classical processors – also opens the door for adaptive programming, whereby the user modifies the program being run based on results at an intermediate stage; this too will require remote access as enabled by a quantum network. Finally, the eventual ability to use a high-speed quantum link to join multiple quantum processors together will greatly enhance their computational capabilities.  For example, two entangled $N$-qubit quantum processors are more powerful, as metricized by the number of bits needed to simulate them on a classical processor, equal to $2^{2N}$, than the two acting independently [70], [71].

Beyond communication and computation applications, quantum entanglement can be used to achieve sensitive object detection and improved (multi-)parameter estimation [45], even in the presence of loss and noise [72]. For example, distributed quantum sensors have the potential to outperform classical devices in metrological tasks like clock synchronization [43], where distributed $N$-party entangled states show a square-root-of-$N$ enhancement to the measurement precision compared to $N$ unentangled clocks, a general scaling improvement for distributed entangled sensors. Similarly, using distributed continuous-variable quantum states, e.g., squeezed states, one can achieve improved sensor performance, though here loss becomes a significant challenge [73], [74].

In a rather different scheme, path-entangled single-photon states distributed to two or more telescopes can be used as nonlocal reference fields (oscillators) that are shot-noise free. The result is that one can use the resulting entangled array of detectors to perform efficient very long baseline interferometric (VLBI) telescopy [75], [76]; if the quantum-entangled states are pre-distributed, *e.g.*, via teleportation across a quantum network, this scheme can circumvent the usual losses or infeasible construction costs that currently limit the telescope separations in optical-range VLBI systems to hundreds of meters [77]. The greatly increased angular resolution achievable with such a telescope array would enable enhanced resolution of astronomical objects in the visible spectrum, including other objects in our solar system.  The same applies to making detailed measurements of Earth systems, assuming one has an array of quantum connected telescopes in space, looking Earthward.

As with all topics discussed in this roadmap, interdisciplinary research and development will be necessary to bring a QTDP focused on applications of quantum entanglement distribution networks to fruition. Different applications of these networks will involve different teams of PIs working together with industry partners and different potential user groups, likely focused on a single goal-oriented application. As such, distinct QTDPs will likely be proposed by different groups within this topic area. A given QTDP will focus on a particular set of technologies needed for that application. Successful groups might emphasize the cross-cutting nature of such applications and strive to unify their activities with other QTDPs in quantum sensing or in local area quantum networking.

### II.A.vi Quantum datacenters

Large-scale high-performance computing in the classical domain was enabled by efficiently networking diverse computational resources and storage devices together in a flexible and modular fashion. Recent advances in modular architectures and software tools allow dynamic allocation of the computational resources by a multitude of users for their needs, a hallmark of modern-day datacenters. Anticipating similar advantage in quantum datacenters, it is crucial for the field to engage in planning for quantum datacenter QTDP as indicated in Fig. 2.

The maturation for individual quantum processing unit capabilities and quantum communication technology will enable modular, distributed quantum computer architectures that will in turn enable scalable quantum datacenters. Like today's classical datacenters and high-performance computing facilities, quantum datacenters will serve as centralized and shared infrastructure that can support high-capacity quantum computational demand from a wide range of users.

A QTDP focused on designing, building and operating a quantum datacenter will require, as a prerequisite, substantial technological progress on (1) individual quantum processing unit technology, (2) local-area quantum networks capable of reliably distributing quantum resources within the datacenter, (3) transduction of quantum data between individual quantum processing units to be communicated over the local-area network, and (4) resource management tools that can adequately manage and allocate various resources in the quantum data center. A QTDP will stimulate and enable new research activities in the areas exploring (a) hybrid quantum computing approaches based on quantum processing units, quantum storage units, and classical resources; (b) novel modular quantum computer architectures leveraging a multitude of quantum processing unit technologies, such as atomic, photonic, and solid-state qubits; (c) development of new quantum algorithms and applications leveraging highly scalable, networked quantum processing units; (d) optimal quantum resource allocation and management methods in a multi-user environment; and (e) formation of a new high-performance quantum computing community (or industry) focused on value creation with future quantum data centers.

A QTDP for quantum datacenters could evolve through several stages of development. In the first 'definition stage,' a consortium of academic researchers and industry partners would define the requirements of a quantum datacenter over different phases of the project, targeting a list of applications that such a datacenter will enable and support. Next is a 'construction stage,' where a QDC will be constructed, operated, attracting a wide range of users implementing various quantum applications. This virtuous cycle of building and productively utilizing generations of QDCs is likely to stimulate a successful industry supporting high-performance quantum computing.

## II.B. Distributed Quantum Technology Hubs

Accomplishing the goals described in this roadmap can be supported by creating distributed quantum technology hubs. These hubs will act as coordinating entities spurring the development, industrialization, and commercialization of quantum technologies, training the future quantum workforce, and building education pipelines that yield diverse participation.

Here, we envision regional hubs that have quantum maker spaces, with expertise and equipment required to fabricate low-volume, cross-cutting, state-of-the-art devices that will vastly increase the quantum toolbox. In these maker spaces, users can work alongside experts to be trained in the fabrication and manufacturing process, mature technologies developed in partnership labs, build commercialization prototypes, and access resources for assisting early-stage quantum startup companies. The proximity between these maker spaces and the industry branches can greatly facilitate convergence of ideas from scientific research to commercial applications. These hubs should be informed by the quantum metrology and certification programs, for example in the US at NIST and the QED-C.

The hubs would also sub-contract to university and industry partners to contribute to their development of key quantum technologies. Currently, startup companies can receive support from the Small Business Innovation Research (SBIR) and Small Business Technology Transfer (STTR) programs (see Sec. III.C below), but in their present forms these programs do not necessarily match with the rapid pace of the quantum information revolution. To more rapidly develop core quantum technologies requested by the QTDPs, the hubs could sub-contract to industry directly. This can include quantum industries and quantum enabling technologies such as vacuum and cryogenic systems, lasers, quantum light sources and detectors, etc. that constitute the quantum technology supply chain. Also, the hubs could advise companies in rapid manufacturing methods and could provide connections to the web of quantum technology users.

These regional centers will also have the possibility of impacting a large swath of the QIST workforce. Because of the centers' distributed nature and their focus on narrower topics, they could lower the barrier for entry into the quantum community. Key pillars of the workforce development within the hubs will be diversity, equity, inclusion, and strengthening and reinforcing a quantum talent pool. These centers could operate or coordinate education and training activities and could implement public outreach programs, including those targeting underrepresented communities, which hold a large reservoir of untapped talent. Also, the hubs will serve as a networking resource with knowledge of the quantum community technology needs, potential market, and connections into larger industries.

## III.  Enhancing the impact of the quantum technology demonstration projects

Finally, in the following we describe other areas we identify that will accelerate achieving a quantum advantage across the sub-disciplines of QIST.

### III.A Long-Range Community Planning, and QTDP Oversight and Metrics

In many collaborative contexts, the QIST community has thrived based on single-investigator and smaller team research activities and this will continue to be supported by the various national programs.  There will be a subset of the community that will want to engage in the larger-scale QTDPs discussed here.  The financial resources supporting the QTDPs will be substantial and necessitate greater accountability to the wider QIST community and to the public.  Thus, it is important to have regular community-wide planning activities that are updated on the few-year time scale, and to develop quantifiable and verifiable metrics that are updated regularly to match the pace of the field.  Metrics and benchmarks have begun to develop in the quantum computing ecosystem [78], and these and similar efforts should be amplified in all areas of QIST.

For example, in the US the original (2018, [79]) and revised (2022, [80]) National Quantum Initiative Act calls out several committees to oversee coordination, strategic planning, and independent assessment of the quantum activities taking place under the Act.  The National Quantum Initiative Coordination Office [81] is responsible for coordination across the government agencies, and joint solicitation and selection of proposals (see the next sub-section on coordination across US government agencies).  The Subcommittee on Quantum Information Science is responsible for fostering the creation of international standards and strategic planning for the community.  Finally, the National Quantum Initiative Advisory Committee (NQIAC) [82] is responsible for reporting on trends in the field, progress toward goals, management and coordination, and implementation of programs, and providing biennial reports containing an independent assessment of activities under the Act.  While some of these crucial activities have been initiated, the sooner they can provide additional output to support the growing QIST community, the more efficiently resources can be applied towards demonstrating a proven quantum advantage.

It may be beneficial for these committees to engage the broader community in their charge by appointing *ad hoc* members that will undertake the initial planning that represents the diverse research activities.  In the US, they might solicit membership through consultation with the National Academies and the professional societies such as Optica, the American Physical Society, IEEE, the Materials Research Society, and from the QED-C, with a similar well-thought-out mix of organizations in other nations.  Furthermore, in the US we suggest that NQIAC be empowered to organize or appoint others to organize town-hall meetings at conferences and workshops that have strong QIST representation.  Such well-balanced and representative national efforts are, in our view, essential to the success of the QTDP endeavor, to ensure that particular interests do not skew investments in a scientifically biased or unhealthy way.

### III.B Case Study: Cooperation across US government agencies

As mentioned, the National Quantum Initiative directs the US government agencies active in supporting QIST research to coordinate their activities.  This has the benefit of avoiding duplication of effort and helping to accelerate achieving quantum advantage across the spectrum of quantum computing, simulation, sensing, and communication.  The quantum networking interagency working group [83] has released a series of technical recommendations (TRs) and programmatic recommendations (PRs) associated with how cooperation across US government agencies can be facilitated.  While these recommendations were specifically for quantum networks, they can be adapted for QIST in general, as QIST systems are typically an integration of quantum sensing, computing, and networking at some level.  The adapted recommendations from the quantum networking interagency working group  relevant to fostering this cooperation are:

TR 1: Continue Research on Use Cases for Quantum Information Systems
TR 2: Prioritize Cross-Beneficial Core Components for Quantum Information Systems
TR 3: Improve Classical Capabilities to Support Quantum Information Systems
TR 4: Leverage "Right-Sized" Quantum Information Systems Testbeds
PR 1: Increase Interagency Coordination on Quantum Information Systems R&D
PR 2: Establish Timetables for Quantum Information Systems R&D Infrastructure

Each agency brings strengths to the overall development of quantum information systems. By leveraging these strengths, synergistic development can be achieved. In one example, a variety of federal agencies have begun the development of a quantum network in the metropolitan area of Washington, DC (DC-QNet) [84].  In another example, several Department of Defense (DoD) agencies are exploring and analyzing the national security implications of advances in quantum technologies, especially in the areas of quantum sensing and quantum networks, while investment in quantum computing includes applications in quantum chemistry, resource optimization, benchmarking, and machine learning. Under Section 214 of the FY2021 National Defense Authorization Act (NDAA) [85], the Secretary of each military department has been tasked to identify "technical problems and research challenges which are likely to be addressable by quantum computers available for use in the next one to three years, with a priority for technical problems and challenges where quantum computing systems have performance advantages over traditional computing systems." Further, the DoD will "establish programs and enter into agreements with appropriate medium and small businesses with functional quantum computing capabilities to provide such private sector capabilities to government, industry, and academic researchers working on relevant technical problems and research activities." A recent DoD quantum information science roadmap acknowledges the magnitude of these challenges and explicitly calls for interagency cooperation to address them.

Collaboration among NSF, DoE, NIST, DoD, and other agencies regarding common or complementary research interests and dissemination of information would greatly facilitate the shared goal of ascertaining and exploiting quantum advantages across the spectrum of quantum computing, simulation, sensing, communication, and  network applications, while simultaneously developing and fostering a quantum engineering workforce through shared faculty/student and academic/government/industrial laboratory fellowship programs.

Additional synergistic benefits will arise from issuing joint solicitations for QTDPs and grants for planning QTDPs involving multiple government agencies, where each agency may elect whether to co-sponsor the entire program or selectively fund the portions of the program most relevant to their mission. For example, one agency, such as the NSF, might fund the initial science-focused project and the Fig. 1 (4 → 5 → 6 → 4) technology-development loop, which could then transition to a different agency at a later stage to address a mission-critical need of the agency. This will also reduce the burden on investigators by lowering the number of proposals submitted and projects managed.

### III.C. Generating Startup Company Support

An important strategy addressed by many nations in the success of QIST is how to support startup companies such as spin-offs from academia or innovative ideas by quantum scientists and engineers striking out into industry, especially newly minted PhDs and post-docs who may not know how to attract venture capital funding. Healthy and successful startups are vital to achieving practical quantum advantage and creating a robust quantum supply chain. Although each nation's system is quite different, nevertheless we believe that useful lessons can be drawn from a careful analysis of the following detailed recommendations relevant to the US system. Especially because venture capital is much less available in some countries, this analysis can be quite useful.

In the US, startup companies can receive support from the Small Business Innovation Research (SBIR) and Small Business Technology Transfer (STTR) programs. The SBIR/STTR grants are important sources of funding for the development of quantum technologies and can help to accelerate the path from laboratory research to commercial deployment and offer an excellent opportunity for businesses to obtain access to non-dilutive seed capital. Therefore, they can be an avenue for enabling companies to pursue higher risk and high-payoff developments expected for quantum technologies.

Industry-university partnerships through the QTDPs, in concert with the SBIR/STTR program, can accelerate the economic impact of QIST. We envision that the QTDPs and quantum technology hubs have an association with companies receiving awards through the SBIR/STTR program so they can gain access to technology and experts in a collaborative way. Also, the companies will have access to a pool of young researchers such as graduate students, post-docs and early-career research scientists that can gain entrepreneurial and project management experience through internships.

However, there are also limitations and barriers that can potentially reduce the effectiveness of these programs, especially for emerging sectors such as QIST. Several obstacles should be kept in mind in developing and utilizing these programs to accelerate innovation in quantum. The size of the awards should be made commensurate with the effort and time required to develop proposals, which will also incentivize the participation of innovators. Additionally, while clear metrics of success toward commercialization are important, it may be beneficial, at least in the arena of quantum technologies, to provide additional metrics/milestones to rank success so that a company may be more willing to participate. Finally, program developers and managers should remain cognizant of the time and effort associated with the administrative burdens imposed by excessive application and reporting requirements for awards to ensure that they do not get in the way of technical progress. Despite these caveats the role of private/public partnerships is vital for the development of QIST, especially to spur creativity and innovation. To that end, the use of these programs in developing quantum technologies can viewed as an opportunity to eliminate challenges, reduce barriers to use, and benchmark the effectiveness of changes in how they are used to create innovation.

### III.D Role of Planning Grants

It is important to start the planning process for standing up new QTDPs such as those indicated in Fig. 2.  Teams of researchers should anticipate, through timed announcements, that funding in specific QTDP focus areas may be available soon. Discussions should start as early as possible utilizing the mechanism of planning grants to form teams with complementary expertise. To maximize the impact, planning grants should include seed funds for well-defined, synergy-building research efforts that both validate and ramp up the targeted QTDP activities. This will not only clarify the roadmap for the actual QTDP but also define concrete research outcomes and impact by which to assess the competing QTDP proposals or teams joining existing QTDPs.

For the longer term and to bring in newly "quantum activated" researchers, planning grants should also have flexibility to allow for newly formed small teams to initiate efforts generating preliminary data that will support or extend existing QTDPs.  Calls for seed proposals should be coordinated and have recurring timelines. Additionally, planning grants should have mechanisms to allow for visiting positions at existing QTDPs, which will foster developing lasting partnerships among universities, industries, and national labs as well as defense laboratories. The visiting positions could take the form of named sabbaticals on the time scale of one-half to one year in duration.

Logistic and organizational mechanisms could be implemented, through a variety of avenues, to facilitate the development of new convergent teams. For example, workshops, team-building exercises, and shark-tank-style pitch opportunities could be organized and supported by national grant agencies such as DOD, DOE, and NSF, industrial consortia such as QED-C, or via steering committees for annual conferences within professional societies such as the Quantum 2.0 and the American Physical Society (APS) meetings. Teaming between large universities and small universities should be encouraged to share expertise and to provide access to equipment necessary for quantum research.

Flexibility in the review process will be needed. Autonomy for funding subcontracts could result in a process that is rapid and agile. Instead of the usual grant review panels drawn almost entirely from academics, a selected technical peer-review group could assess planning grants and provide recommendations to an advisory board consisting of members from government, academia, and industry for final selection. Support for teaming using this non-peer-reviewed approach could help to accelerate timelines for research progress and maintain research focus with the trajectory of the QTDPs.

## IV. Conclusions and Outlook

In this Roadmap we have proposed a necessary step in bridging the technology readiness levels 1-3 found mainly at universities to TRLs 7-9 typical of successful industry and public access efforts.  This bridge, focused on TRLs 4-6, encompasses Quantum Technology Demonstration Projects (QTDPs) – large-scale public-private partnerships that have high probability for translation from laboratory to practice, creating technology that demonstrates a clear 'quantum advantage' for science breakthroughs that is user-motivated and will provide access to a broad and diverse community of scientific users. Successful implementation of a program of QTDPs will have large positive economic impacts.  We provide a concrete priority list of six suggested QTDPs in our roadmap, together with clear technical justification and explanation.

Finally, there are several other topics not appearing in this roadmap that are not yet at the level for standing up a QTDP.  These should continue to be supported by existing fundamental research mechanisms, such as the NSF Quantum Leap Challenge Institutes in the US. Importantly, multiple efforts should be supported that are aimed directly at removing identified bottlenecks. For example, high-rate and high-efficiency quantum repeaters for long-distance entanglement distribution are still missing, although many concepts for building such are known, and could be pursued by teams that are aiming to stand up a QTDP in long-distance networking.

## Acknowledgements

We gratefully acknowledge the financial support of the National Science Foundation under grant No. PHY-2230199 and the travel and event support of MaLinda Hill of The Ohio State University Enterprise for Research, Innovation, and Knowledge and Melanie Holbert of The Ohio State University Department of Physics.

This work is based upon a Project Scoping Workshop, which took place June 27-28, 2022 at The Ohio State University in Columbus, OH, USA, where approximately forty representative members of the United States of America QIST community carried out a scoping exercise to address the following primary questions:

- What topics in QIST are now ready or near-ready for a major push via large-scale public-private partnerships that have high probability to lead to translation from laboratory to practice, creating technology that demonstrates a clear 'quantum advantage' for science breakthroughs that is user-motivated and will provide access to a broad and diverse community of scientific users throughout the country?
- What mechanisms should be put in place to accelerate the large-scale projects?

The workshop participants included the needed range of expertise and experience: leading researchers at universities, including faculty at historically black colleges and universities (HBCUs); industry and government civilian and defense laboratories; representatives of companies that collaborate and/or develop and supply critical products to these researchers; and representatives of industry organizations.